\PassOptionsToPackage{unicode}{hyperref}
\PassOptionsToPackage{hyphens}{url}
\PassOptionsToPackage{dvipsnames,svgnames,x11names}{xcolor}
\documentclass[
]{article}
\usepackage{amsmath,amssymb}
\usepackage{lmodern}
\usepackage{iftex}
\ifPDFTeX
  \usepackage[T1]{fontenc}
  \usepackage[utf8]{inputenc}
  \usepackage{textcomp} 
\else 
  \usepackage{unicode-math}
  \defaultfontfeatures{Scale=MatchLowercase}
  \defaultfontfeatures[\rmfamily]{Ligatures=TeX,Scale=1}
\fi
\IfFileExists{upquote.sty}{\usepackage{upquote}}{}
\IfFileExists{microtype.sty}{
  \usepackage[]{microtype}
  \UseMicrotypeSet[protrusion]{basicmath} 
}{}
\makeatletter
\@ifundefined{KOMAClassName}{
  \IfFileExists{parskip.sty}{%
    \usepackage{parskip}
  }{
    \setlength{\parindent}{0pt}
    \setlength{\parskip}{6pt plus 2pt minus 1pt}}
}{
  \KOMAoptions{parskip=half}}
\makeatother
\usepackage{xcolor}
\providecommand{\tightlist}{%
  \setlength{\itemsep}{0pt}\setlength{\parskip}{0pt}}
\newlength{\cslhangindent}
\newlength{\csllabelwidth}
\newlength{\cslentryspacingunit} 
\newenvironment{CSLReferences}[2] 
 {
  \setlength{\parindent}{0pt}
  \ifodd #1
  \let\oldpar\par
  \def\par{\hangindent=\cslhangindent\oldpar}
  \fi
  \setlength{\parskip}{#2\cslentryspacingunit}
 }%
 {}
\usepackage{calc}

\ifLuaTeX
\usepackage[bidi=basic]{babel}
\else
\usepackage[bidi=default]{babel}
\fi
\babelprovide[main,import]{american}

\def\languageshorthands#1{}
\ifLuaTeX
  \usepackage{selnolig}  
\fi
\IfFileExists{bookmark.sty}{\usepackage{bookmark}}{\usepackage{hyperref}}
\IfFileExists{xurl.sty}{\usepackage{xurl}}{} 
\hypersetup{
  pdftitle={Paicos: A Python package for analysis of (cosmological)
simulations performed with Arepo},
  pdfauthor={Thomas Berlok, Léna Jlassi, Ewald Puchwein, Troels
Haugbølle},
  pdflang={en-US},
  colorlinks=true,
  linkcolor={Maroon},
  filecolor={Maroon},
  citecolor={Blue},
  urlcolor={Blue},
  pdfcreator={LaTeX via pandoc}}

\title{Paicos: A Python package for analysis of (cosmological)
simulations performed with Arepo}

\usepackage[affil-it]{authblk}
\usepackage{orcidlink}
\author[1,2%
  \ensuremath\mathparagraph]{Thomas Berlok%
    \,\orcidlink{0000-0003-0466-603X}\,%
    }
\author[2%
  ]{Léna Jlassi%
    \,\orcidlink{0009-0007-9039-294X}\,%
    }
\author[2%
  ]{Ewald Puchwein%
    \,\orcidlink{0000-0001-8778-7587}\,%
    }
\author[1%
  ]{Troels Haugbølle%
    \,\orcidlink{0000-0002-9422-8684}\,%
    }

\affil[1]{Niels Bohr Institute, University of Copenhagen, Denmark}
\affil[2]{Leibniz-Institut für Astrophysik Potsdam, Germany}
\affil[$\mathparagraph$]{Corresponding author}
\date{31 January 2024}

\begin{document}
\maketitle

\hypertarget{summary}{%
\section{Summary}\label{summary}}

Cosmological simulations evolve dark matter and baryons subject to
gravitational and hydrodynamic forces
(\protect\hyperlink{ref-Vogelsberger2020}{Vogelsberger et al., 2020}).
The simulations start at high redshift and capture hierarchical
structure formation where small structures form first and later assemble
to larger structures (\protect\hyperlink{ref-Springel2005}{Springel et
al., 2005}). The Arepo code is a versatile finite-volume code which can
solve the magnetohydrodynamic equations on an unstructured Voronoi mesh
in a cosmologically comoving frame
(\protect\hyperlink{ref-Springel2010}{Springel, 2010};
\protect\hyperlink{ref-Weinberger2020}{Weinberger et al., 2020}).

Here we present Paicos, a new object-oriented Python package for
analyzing simulations performed with Arepo. Paicos strives to reduce the
learning curve for students and researchers getting started with Arepo
simulations. As such, Paicos includes many examples in the form of
Python scripts and Jupyter notebooks
(\protect\hyperlink{ref-Kluyver2016}{Kluyver et al., 2016}) as well as
an online documentation describing the installation procedure and
recommended first steps.

Paicos' main features are automatic handling of cosmological and
physical units, computation of derived variables, 2D visualization
(slices and projections), 1D and 2D histograms, and easy saving and
loading of derived data including units and all the relevant metadata.

Paicos relies heavily on well-established open source software libraries
such as NumPy (\protect\hyperlink{ref-Harris2020}{Harris et al., 2020};
\protect\hyperlink{ref-numpy}{van der Walt et al., 2011}), Scipy
(\protect\hyperlink{ref-Virtanen2020}{Virtanen et al., 2020}), h5py
(\protect\hyperlink{ref-h5py}{Collette, 2013}), Cython
(\protect\hyperlink{ref-Behnel2011}{Behnel et al., 2011}) and astropy
(\protect\hyperlink{ref-astropy}{Astropy Collaboration, 2013}) and
contains features for interactive data analysis inside Ipython terminals
(\protect\hyperlink{ref-Perez2007}{Perez \& Granger, 2007}) and Jupyter
notebooks (\protect\hyperlink{ref-Kluyver2016}{Kluyver et al., 2016}),
e.g., tab completion of data keywords and Latex rendering of data units.
Paicos also contains a number of tests that are automated using pytest
(\protect\hyperlink{ref-pytest7.4}{Krekel et al., 2004}), CircleCi and
GitHub workflows.

\hypertarget{statement-of-need}{%
\section{Statement of need}\label{statement-of-need}}

The Arepo code stores its data output as HDF5 files, which can easily be
loaded as NumPy arrays using h5py. However, data visualization of the
unstructured mesh used in Arepo is non-trivial and keeping track of the
units used in the data outputs can also be a tedious task. The general
purpose visualization and data analysis software package yt\footnote{See
  http://yt-project.org/.} (\protect\hyperlink{ref-yt-Turk}{Turk et al.,
2011}) and the visualization package py-sphviewer
(\protect\hyperlink{ref-py-sphviewer}{Benitez-Llambay, 2015}) are both
able to perform visualizations of Arepo simulations. Paicos provides an
alternative, which is specifically tailored to Arepo simulations. It is
in this regard similar to swiftsimio
(\protect\hyperlink{ref-Borrow2020}{Borrow \& Borrisov, 2020}), which
was developed specifically for SWIFT simulations
(\protect\hyperlink{ref-Schaller2023}{Schaller \& others, 2023}).

We have developed Paicos because we identified a need for an analysis
code that simultaneously fulfills the following requirements: 1) is
specifically written for analysis of Arepo simulations discretized on a
Voronoi mesh 2) provides safeguards against errors related to
conversions of cosmological and physical units 3) facilitates working
with large data sets by supporting the saving and loading of
reduced/derived data including units and metadata 4) contains enough
functionality to be useful for practical research tasks while still
being light-weight and well-documented enough that it can be installed,
used and understood by a junior researcher with little or no assistance.

\hypertarget{overview-of-key-paicos-features}{%
\section{Overview of key Paicos
features}\label{overview-of-key-paicos-features}}

The key Paicos features, which were implemented to fulfill the above
requirements, are as follows:

\begin{itemize}
\tightlist
\item
  Functionality for reading cell/particle and group/subhalo catalog data
  as saved by Arepo in the HDF5 format.
\item
  Unit-handling via astropy with additional support for the \(a\) and
  \(h\) factors that are used in cosmological simulations.\footnote{See
    e.g. Pakmor \& Springel (\protect\hyperlink{ref-Pakmor2013}{2013})
    and Berlok (\protect\hyperlink{ref-Berlok2022}{2022}) for detailed
    discussions and a derivation of the comoving magnetohydrodynamic
    equations used in cosmological simulations including magnetic
    fields.}
\item
  Functionality for automatically obtaining derived variables from the
  variables present in the Arepo snapshot (e.g.~the cell volume,
  temperature, or magnetic field strength). Within Jupyter notebooks or
  an Ipython terminal: Tab completion to show which variables are
  available for a given data group (either by directly loading from the
  HDF5-file or by automatically calculating a derived variable).
\item
  Functionality for saving and loading data including cosmological and
  physical units in HDF5-files that automatically include all metadata
  from the original Arepo snapshot (i.e.~the Header, Config and Param
  groups).
\item
  Functionality for creating slices and projections and saving them in a
  format that additionally includes information about the size, center
  and orientation of the image.
\item
  Functionality for creating 1D and 2D histograms (e.g.~radial profiles
  and \(\rho\)-\(T\) phase-space plots) of large data sets (using
  OpenMP-enabled Cython).
\item
  Functionality for user customization, e.g.~adding new units or
  functions for computing derived variables.
\item
  Functionality for selecting only a part of a snapshot for analysis and
  for storing this selection as a new reduced snapshot.
\end{itemize}

Finally, we also make public GPU implementations of projection
functionalities, which are much faster than the OpenMP-parallel CPU
implementations described above. The GPU implementations include a
mass-conserving SPH-like projection and a ray tracing implementation
using a bounding volume hierarchy (BVH) in the form of a binary radix
tree for nearest neighbor searches. These GPU implementations use Numba
(\protect\hyperlink{ref-Numba2015}{Lam et al., 2015}) and CuPy
(\protect\hyperlink{ref-cupy_learningsys2017}{Okuta et al., 2017}) and
require that the user has a CUDA-enabled GPU. Our binary tree
implementation follows the the GPU-optimized tree-construction algorithm
described in (\protect\hyperlink{ref-Karras2012}{Karras, 2012}) and the
implementation of it found in the publicly available Cornerstone Octree
GPU-library (\protect\hyperlink{ref-Keller2023}{Keller et al., 2023}).
Using our SPH-like implementation on an
\href{https://www.nvidia.com/en-us/data-center/a100/}{Nvidia A100 GPU},
it takes less than a second to project 100 mio. particles onto an image
plane with \(4096^2\) pixels. This speed enables interactive data
exploration. We provide an example IPython widget illustrating this
feature. Finally, we note that the returned images include physical
units and can be used for scientific analysis.

Paicos is hosted on GitHub at https://github.com/tberlok/paicos. We
strongly encourage contributions to Paicos by opening issues and
submitting pull requests.

\hypertarget{acknowledgements}{%
\section{Acknowledgements}\label{acknowledgements}}

We thank the JOSS referees, Kyle Oman and Terrence Tricco, and the
editor, Josh Borrow, for constructive feedback that greatly helped
improve Paicos. We are grateful to Christoph Pfrommer for support and
advice. We thank Rüdiger Pakmor, Rosie Talbot and Timon Thomas for
useful discussions as well as Matthias Weber, Lorenzo Maria Perrone,
Arne Trabert and Joseph Whittingham for beta-testing Paicos. We are
grateful to Volker Springel for making the Arepo code available and to
the main developers of arepo-snap-util (Federico Marinacci and Rüdiger
Pakmor), a non-public code which we have used for comparison of
projections and slices. TB gratefully acknowledges funding from the
European Union's Horizon Europe research and innovation programme under
the Marie Skłodowska-Curie grant agreement No 101106080. LJ acknowledges
support by the German Science Foundation (DFG) under grant ``DFG
Research Unit FOR 5195 -- Relativistic Jets in Active Galaxies''. LJ and
TB acknowledge support by the European Research Council under ERC-AdG
grant PICOGAL-101019746. The authors gratefully acknowledge the Gauss
Centre for Supercomputing e.V. (www.gauss-centre.eu) for funding this
project by providing computing time on the GCS Supercomputer SuperMUC at
Leibniz Supercomputing Centre (www.lrz.de). The Tycho supercomputer
hosted at the SCIENCE HPC center at the University of Copenhagen was
used for supporting this work.

\hypertarget{references}{%
\section*{References}\label{references}}
\addcontentsline{toc}{section}{References}

\hypertarget{refs}{}
\begin{CSLReferences}{1}{0}
\leavevmode\vadjust pre{\hypertarget{ref-astropy}{}}%
Astropy Collaboration. (2013). {Astropy: A community Python package for
astronomy}. \emph{Astronomy and Astrophysics}, \emph{558}.
\url{https://doi.org/10.1051/0004-6361/201322068}

\leavevmode\vadjust pre{\hypertarget{ref-Behnel2011}{}}%
Behnel, S., Bradshaw, R., Citro, C., Dalcin, L., Seljebotn, D. S., \&
Smith, K. (2011). {Cython: The Best of Both Worlds}. \emph{Computing in
Science and Engineering}, \emph{13}(2), 31--39.
\url{https://doi.org/10.1109/MCSE.2010.118}

\leavevmode\vadjust pre{\hypertarget{ref-py-sphviewer}{}}%
Benitez-Llambay, A. (2015). \emph{Py-sphviewer: Py-SPHViewer v1.0.0}
(Version v1.0.0). Zenodo. \url{https://doi.org/10.5281/zenodo.21703}

\leavevmode\vadjust pre{\hypertarget{ref-Berlok2022}{}}%
Berlok, T. (2022). {Hydromagnetic waves in an expanding universe -
cosmological MHD code tests using analytic solutions}. \emph{Monthly
Notices of the RAS}, \emph{515}(3), 3492--3511.
\url{https://doi.org/10.1093/mnras/stac1882}

\leavevmode\vadjust pre{\hypertarget{ref-Borrow2020}{}}%
Borrow, J., \& Borrisov, A. (2020). Swiftsimio: A python library for
reading SWIFT data. \emph{Journal of Open Source Software},
\emph{5}(52), 2430. \url{https://doi.org/10.21105/joss.02430}

\leavevmode\vadjust pre{\hypertarget{ref-h5py}{}}%
Collette, A. (2013). \emph{Python and HDF5}. O'Reilly.

\leavevmode\vadjust pre{\hypertarget{ref-Harris2020}{}}%
Harris, C. R., Millman, K. J., van der Walt, S. J., Gommers, R.,
Virtanen, P., Cournapeau, D., Wieser, E., Taylor, J., Berg, S., Smith,
N. J., Kern, R., Picus, M., Hoyer, S., van Kerkwijk, M. H., Brett, M.,
Haldane, A., del Rio, J. F., Wiebe, M., Peterson, P., \ldots{} Oliphant,
T. E. (2020). {Array programming with NumPy}. \emph{Nature},
\emph{585}(7825), 357--362.
\url{https://doi.org/10.1038/s41586-020-2649-2}

\leavevmode\vadjust pre{\hypertarget{ref-Karras2012}{}}%
Karras, T. (2012). Maximizing parallelism in the construction of BVHs,
octrees, and k-d trees. \emph{Proceedings of the Fourth ACM SIGGRAPH /
Eurographics Conference on High-Performance Graphics}, 33--37.
\url{https://doi.org/10.2312/EGGH/HPG12/033-037}

\leavevmode\vadjust pre{\hypertarget{ref-Keller2023}{}}%
Keller, S., Cavelan, A., Cabezon, R., Mayer, L., \& Ciorba, F. M.
(2023). {Cornerstone: Octree Construction Algorithms for Scalable
Particle Simulations}. \emph{arXiv e-Prints}, arXiv:2307.06345.
\url{https://doi.org/10.48550/arXiv.2307.06345}

\leavevmode\vadjust pre{\hypertarget{ref-Kluyver2016}{}}%
Kluyver, T., Ragan-Kelley, B., Pérez, F., Granger, B., Bussonnier, M.,
Frederic, J., Kelley, K., Hamrick, J., Grout, J., Corlay, S., Ivanov,
P., Avila, D., Abdalla, S., Willing, C., \& team, J. development.
(2016). Jupyter notebooks -- a publishing format for reproducible
computational workflows. In F. Loizides \& B. Scmidt (Eds.),
\emph{Positioning and power in academic publishing: Players, agents and
agendas} (pp. 87--90). IOS Press.
\url{https://doi.org/10.3233/978-1-61499-649-1-87}

\leavevmode\vadjust pre{\hypertarget{ref-pytest7.4}{}}%
Krekel, H., Oliveira, B., Pfannschmidt, R., Bruynooghe, F., Laugher, B.,
\& Bruhin, F. (2004). \emph{Pytest 7.4}.
\url{https://github.com/pytest-dev/pytest}

\leavevmode\vadjust pre{\hypertarget{ref-Numba2015}{}}%
Lam, S. K., Pitrou, A., \& Seibert, S. (2015). Numba: A LLVM-based
python JIT compiler. \emph{Proceedings of the Second Workshop on the
LLVM Compiler Infrastructure in HPC}.
\url{https://doi.org/10.1145/2833157.2833162}

\leavevmode\vadjust pre{\hypertarget{ref-cupy_learningsys2017}{}}%
Okuta, R., Unno, Y., Nishino, D., Hido, S., \& Loomis, C. (2017). CuPy:
A NumPy-compatible library for NVIDIA GPU calculations.
\emph{Proceedings of Workshop on Machine Learning Systems (LearningSys)
in the Thirty-First Annual Conference on Neural Information Processing
Systems (NIPS)}.
\url{http://learningsys.org/nips17/assets/papers/paper_16.pdf}

\leavevmode\vadjust pre{\hypertarget{ref-Pakmor2013}{}}%
Pakmor, R., \& Springel, V. (2013). {Simulations of magnetic fields in
isolated disc galaxies}. \emph{Monthly Notices of the RAS},
\emph{432}(1), 176--193. \url{https://doi.org/10.1093/mnras/stt428}

\leavevmode\vadjust pre{\hypertarget{ref-Perez2007}{}}%
Perez, F., \& Granger, B. E. (2007). {IPython: A System for Interactive
Scientific Computing}. \emph{Computing in Science and Engineering},
\emph{9}(3), 21--29. \url{https://doi.org/10.1109/MCSE.2007.53}

\leavevmode\vadjust pre{\hypertarget{ref-Schaller2023}{}}%
Schaller, M., \& others. (2023). {Swift: A modern highly-parallel
gravity and smoothed particle hydrodynamics solver for astrophysical and
cosmological applications}. \emph{arXiv e-Prints}, arXiv:2305.13380.
\url{https://doi.org/10.48550/arXiv.2305.13380}

\leavevmode\vadjust pre{\hypertarget{ref-Springel2010}{}}%
Springel, V. (2010). {E pur si muove: Galilean-invariant cosmological
hydrodynamical simulations on a moving mesh}. \emph{Monthly Notices of
the RAS}, \emph{401}(2), 791--851.
\url{https://doi.org/10.1111/j.1365-2966.2009.15715.x}

\leavevmode\vadjust pre{\hypertarget{ref-Springel2005}{}}%
Springel, V., White, S. D. M., Jenkins, A., Frenk, C. S., Yoshida, N.,
Gao, L., Navarro, J., Thacker, R., Croton, D., Helly, J., Peacock, J.
A., Cole, S., Thomas, P., Couchman, H., Evrard, A., Colberg, J., \&
Pearce, F. (2005). {Simulations of the formation, evolution and
clustering of galaxies and quasars}. \emph{Nature}, \emph{435}(7042),
629--636. \url{https://doi.org/10.1038/nature03597}

\leavevmode\vadjust pre{\hypertarget{ref-yt-Turk}{}}%
Turk, M. J., Smith, B. D., Oishi, J. S., Skory, S., Skillman, S. W.,
Abel, T., \& Norman, M. L. (2011). {yt: A Multi-code Analysis Toolkit
for Astrophysical Simulation Data}. \emph{The Astrophysical Journal
Supplement Series}, \emph{192}, 9.
\url{https://doi.org/10.1088/0067-0049/192/1/9}

\leavevmode\vadjust pre{\hypertarget{ref-numpy}{}}%
van der Walt, S., Colbert, S. C., \& Varoquaux, G. (2011). {The NumPy
Array: A Structure for Efficient Numerical Computation}. \emph{Computing
in Science and Engineering}, \emph{13}(2), 22--30.
\url{https://doi.org/10.1109/MCSE.2011.37}

\leavevmode\vadjust pre{\hypertarget{ref-Virtanen2020}{}}%
Virtanen, P., Gommers, R., Oliphant, T. E., Haberland, M., Reddy, T.,
Cournapeau, D., Burovski, E., Peterson, P., Weckesser, W., Bright, J.,
van der Walt, S. J., Brett, M., Wilson, J., Millman, K. J., Mayorov, N.,
Nelson, A. R. J., Jones, E., Kern, R., Larson, E., \ldots{} SciPy 1. 0
Contributors. (2020). {SciPy 1.0: fundamental algorithms for scientific
computing in Python}. \emph{Nature Methods}, \emph{17}, 261--272.
\url{https://doi.org/10.1038/s41592-019-0686-2}

\leavevmode\vadjust pre{\hypertarget{ref-Vogelsberger2020}{}}%
Vogelsberger, M., Marinacci, F., Torrey, P., \& Puchwein, E. (2020).
{Cosmological simulations of galaxy formation}. \emph{Nature Reviews
Physics}, \emph{2}(1), 42--66.
\url{https://doi.org/10.1038/s42254-019-0127-2}

\leavevmode\vadjust pre{\hypertarget{ref-Weinberger2020}{}}%
Weinberger, R., Springel, V., \& Pakmor, R. (2020). {The AREPO Public
Code Release}. \emph{Astrophysical Journal, Supplement}, \emph{248}(2),
32. \url{https://doi.org/10.3847/1538-4365/ab908c}

\end{CSLReferences}

\end{document}